\documentclass[aps,pra,twocolumn,superscriptaddress,amssymb,showpacs]{revtex4}

\usepackage{graphicx}

\begin{document}

%\preprint{}

\title{Quadrature entanglement and photon-number correlations accompanied by phase-locking}



\author{H.~H.~Adamyan}
\email[]{adam@unicad.am}
%\thanks{}
\affiliation{Institute for Physical Research, National Academy of
Sciences,\\Ashtarak-2, 378410, Armenia}

\author{N.~H.~Adamyan}
\email[]{Narek.Adamyan@synopsys.com}
%\thanks{}
\affiliation{Yerevan State University, A. Manookyan 1, 375049,
Yerevan, Armenia}

\author{S.~B.~Manvelyan}
\email[]{smanvel@server.physdep.r.am}
%\thanks{}
\affiliation{Institute for Physical Research, National Academy of
Sciences,\\Ashtarak-2, 378410, Armenia}

\author{G.~Yu.~Kryuchkyan}
\email[]{gkryuchk@server.physdep.r.am}
%\thanks{}
\affiliation{Yerevan State University, A. Manookyan 1, 375049,
Yerevan, Armenia} \affiliation{Institute for Physical Research,
National Academy of Sciences,\\Ashtarak-2, 378410, Armenia}


\begin{abstract}
We investigate quantum properties of phase-locked light beams
generated in a nondegenerate optical parametric oscillator (NOPO)
with an intracavity waveplate. This investigation continuous our
previous analysis presented in Phys.Rev.A \textbf{69}, 05814
(2004), and involves problems of continuous-variable quadrature
entanglement in the spectral domain, photon-number correlations as
well as the signatures of phase-locking in the Wigner function. We
study the role of phase-localizing processes on the quantum
correlation effects. The peculiarities of phase-locked NOPO in the
self-pulsing instability operational regime are also cleared up.
The results are obtained in both the $P$-representation as a
quantum-mechanical calculation in the framework of stochastic
equations of motion, and also by using numerical simulation based
on the method of quantum state diffusion.
\end{abstract}

%\keywords{}

\maketitle

\section{INTRODUCTION}

In the presently very active field of continuous variable (CV)
quantum information processing a challenging goal consists in
generation of entangled intensive light beams. Various quantum
optical schemes generating entangled bright light have been
proposed for this goal. For the first time the CV entangled states
of light have been studied in \cite{Reid} and demonstrated
experimentally in \cite{Ou} for nondegenerate optical parametric
oscillator (NOPO) below the threshold. An experimental progress in
this direction has been recently reported in
\cite{Braunstein,Zhang}. However, up to now the generation of
bright light beams with high level of CV entanglement meets
serious problems. One of these is degradation of entanglement due
to uncontrolled dissipation and decoherence, and cavity induced
feedback. Beside this, it is known, that the relative phase
between subharmonics in above threshold NOPO undergoes diffusion
process. This process destroys the frequency degeneracy of modes
and hence limits the production of quantum-twin beams in NOPO
above threshold. For reducing such phase diffusion phenomena
various methods, based on phase locking mechanisms have been
proposed \cite{mason,fabre,Boller,muradyan,zondy,kry}.

The simplest scheme proposed and experimentally realized by Mason
and Wong \cite{mason} is the NOPO with additional intracavity
quarter-wave plate to provide polarization mixing between two
orthogonally polarized modes of the subharmonics. Indeed, it was
shown that induced linear coupling between generated modes results
in a locking phenomenon and hence in frequency degenerate
operation above threshold. Following this experiment, the
semiclassical theory of NOPO with self-phase locking was developed
in \cite{fabre} and its further more detailed consideration has
been presented by Fabre and colleagues \cite{longchambon}. In a
more general approach the semiclassical theory of the
self-phase-locked NOPO has been investigated by Gro$\beta$ and
Boller \cite{Boller}.

Type-II optical parametric oscillators with self-phase licking
have recently attracted a lot of attention as efficient source of
nonclassical light \cite{adam1, adam2,laurat1,laurat2,laurat4,
feng}. A full quantum mechanical treatment of this system in
application to generation of CV entangled states of light beams
under mode phase-locked condition has been presented in
\cite{adam1, adam2}, where the regimes below, near and above
threshold was considered. Quantum optical effects have been
demonstrated by Fabre group in the series of experiments
\cite{laurat1,laurat2,laurat4}. Generation of quadrature EPR
entanglement has been demonstrated experimentally \cite{laurat1}
with self-phase-locked NOPO below threshold. Experimental
investigation of intensity quantum correlation with phase-locked
NOPO above threshold has been also performed \cite{laurat2}. The
intensity-difference squeezing in electronically phase locked NOPO
above threshold as well as the Hong-Ou-Mandel interferometry using
twin beams have also been experimentally demonstrated by Feng and
Pfister \cite{feng}. Stable generation in application to metrology
has been also demonstrated \cite{gross2}.

In this paper we continue investigation of phase-locked NOPO
follow the paper \cite{adam1}. The previous studies of
entanglement in the presence of phase localizing processes have
only concentrated on stable stationary solutions for dynamics of
intracavity subharmonics. Quite recently, the experimental
observation of dynamical signatures of self-phase locking in a
triply resonant degenerate OPO was reported in \cite{zondy1} for
both stable and unstable regimes, however, the study of dynamical
aspects of entanglement for this system have been postponed for
future investigations. In addition, here we will consider
entanglement production in the unstable regime of generation. We
shall also give below the quantum description of the unstable
regime in phase-space on the framework of Wigner function.

Our goal is two-fold. In one part of the present paper, we expand
the previous study of phase-locked NOPO in the stable,
steady-state regime of generation \cite{adam1}, particularly
considering both the important details in phase-space in the frame
of Wigner function and photon-numbers difference squeezing in the
presence of phase locking. We also investigate quadrature EPR
entanglement in the spectral domain in addition to our previous
results which have been performed in the time domain. The other
part of the paper is devoted to the problem of entanglement in the
self-pulsing instability regime.

The system under consideration is the NOPO with quarter wave plate
inside the cavity, interacting with the thermal bath. Due to
explicit presence of dissipation in this problem, one has to write
the master equation for the reduced density matrix of the system,
which within the framework of the rotating wave approximation and
in the interaction picture is
\begin{eqnarray}
\frac{\partial \rho }{\partial t}&=&\frac{1}{i\hbar }\left[H,\rho %
\right]\nonumber\\
&+&\sum_{i=1}^{3}\gamma _{i}\left( 2a_{i}\rho
a_{i}^{+}-a_{i}^{+}a_{i}\rho -\rho a_{i}^{+}a_{i}\right),
\label{MasterEq}
\end{eqnarray}
where
\begin{eqnarray}
H&=&\sum_{i=1}^{3}\hbar \Delta _{i}a_{i}^{+}a_{i}+{i\hbar}E\left(a_{3}^{+}-a_{3}\right)\nonumber \\
&+&i\hbar k (a_{3}a_{1}^{+}a_{2}^{+}-a_{3}^{+}a_{1}a_{2}) +\hbar
\chi(a_{1}^{+}a_{2}+a_{1}a_{2}^{+}), \label{TransformedH}
\end{eqnarray}
where $a_{i}$ are the boson operators for the cavity modes
$\omega_{i}$. The mode $a_{3}$ at frequency $\omega $ is driven by
an external field with amplitude $E$, while $a_{1}$ and $a_{2}$
describe subharmonics of two orthogonal polarizations at the
degenerate frequency $\omega /2$. The constant $k$ determines the
efficiency of the down-conversion process, while $\chi$ describes
the energy exchange between only the subharmonic modes due to the
intracavity waveplate. It should be noted that maximally reached
value for $\chi$ in the experiments is $\chi/\gamma\simeq0.21$.
However in the following calculations we consider $\chi/\gamma$ in
a more wide range keeping in mind future experimental
achievements. We take into account the detunings of subharmonics
$\Delta _{i}$ and the cavity damping rates $\gamma _{i}$ and
consider the case of high cavity losses for pump mode ($\gamma
_{3}\gg \gamma_{1}, \gamma_{2}$), when this mode can be
adiabatically eliminated. However, in our analysis we take into
account the pump depletion effects. We will solve
Eq.(\ref{MasterEq}) in the framework of $P$-representation and
stochastic variables on one side, and also by using the well known
numerical quantum state diffusion method (QSD), according to
which, open quantum systems are represented by the ensemble of
quantum trajectories \cite{percival}. The density matrix is
restored in this case from the ensemble averaging over photonic
Fock states.

The paper is arranged as follows. In Sec. II we recall some
theoretical points of self-phase locked NOPO and we present
analysis of both photon-number quantum correlation and phase
locking in the base of the Wigner functions of generated modes.
Section III is devoted to an analysis of quantum fluctuations of
both modes in below-threshold operational regime as well as to
calculation of the squeezed quadrature variance. In Sec. IV  we
investigate the self-pulsing instability regime of NOPO on the
base of quantum trajectories and in the phase space. We also
discuss there the CV entanglement in the unstable regime of
generation. We summarize our results in Sec.V.

\section{Signatures of phase-locking in the quantum correlations and in the phase space}

At first, we shortly discuss the phase-locking phenomena in the
semiclassical theory of NOPO. The stochastic equations of
self-locked NOPO for the complex c-number variables $\alpha _{i}$
and $\beta _{i}$ corresponding to the operators $a_{i}$ and
$a_{i}^{+}$, in the regime of adiabatic elimination of pump mode
have the following form \cite{adam1}:
\begin{eqnarray}
\frac{\partial \alpha _{1}}{\partial t}&=&-\left(\gamma
_{1}+i\Delta
_{1}\right)\alpha_{1}\nonumber \\
&&+\left(\varepsilon -\lambda \alpha _{1}\alpha _{2}\right)\beta
_{2}-i\chi \alpha _{2}+R_{1}, \label{a1StochEq}
\end{eqnarray}
\begin{eqnarray}
\frac{\partial \beta _{1}}{\partial t}&=&-\left( \gamma
_{1}-i\Delta _{1}\right) \beta _{1}\nonumber \\
&&+\left( \varepsilon -\lambda \beta _{1}\beta _{2}\right) \alpha
_{2}+i\chi \beta _{2}+R_{1}^{\dag}. \label{b1StochEq}
\end{eqnarray}
Here $\varepsilon =kE/\gamma _{3}$ and $\lambda =k^{2}/\gamma
_{3}$. The equations for $\alpha _{2}$ and $\beta _{2}$ are
obtained by changing the indexes $(1)\leftrightarrows(2)$.
$R_{1}$, $R_{2}$ are Gaussian noise terms obeying the following
correlations

\begin{eqnarray}
\langle
R_{1}(t)R_{2}(t')\rangle=\frac{k}{\gamma_{3}}(E-k\alpha_{1}\alpha_{2})\delta(t-t'),
\label{correlations1}
\end{eqnarray}
\begin{eqnarray}
\langle
R_{1}^\dag(t)R_{2}^\dag(t')\rangle=\frac{k}{\gamma_{3}}(E-k\beta_{1}\beta_{2})\delta(t-t').
\label{correlations2}
\end{eqnarray}

It has been shown in \cite{longchambon, adam1} that these
equations in the semiclassical approximation  have stationary
stable solutions only if the following relation holds
\begin{equation}
4\chi ^{2}\Delta _{1}\Delta _{2}>\left( \gamma _{1}\Delta
_{2}-\gamma _{2}\Delta _{1}\right) ^{2}, \label{SolutionCondition}
\end{equation}
otherwise the system has only unstable solutions.

It should be noted that the phase locking condition
(\ref{SolutionCondition}) obtained here in the adiabatic
approximation of triply resonant NOPO has a universal form and is
realized for the various schemes of self-phase-locked NOPO. It has
been obtained at first \cite{mason,fabre} for the case of doubly
resonant NOPO. As shown in \cite{Boller} this condition coincides
with analogous condition obtained for triply resonant NOPO without
adiabatic approximation.

Let us recall the results concerned to the phases and photon
numbers for the stationary, stable regime of self locked NOPO for
the cases of $\Delta_{1}=\Delta_{2}\equiv\Delta$, and
$\gamma_{1}=\gamma_{2}=\gamma$ \cite{adam1}. In above threshold
regime
$\varepsilon\geq\varepsilon_{th}=\sqrt{(\chi-|\Delta|)^{2}+\gamma^{2}}$,
the steady-state solution for the photon numbers is:

\begin{equation}
n_{0}=n_{10}=n_{20}=\frac{1}{\lambda }\left[ \sqrt{\varepsilon
^{2}-\left( \chi -\left| \Delta \right| \right) ^{2}}-\gamma
\right], \label{Phot12Simplified}
\end{equation}
while the phases have been found to be

\begin{equation}
\varphi _{10}=\varphi _{20}=-\frac{1}{2}Arc\sin
\frac{1}{\varepsilon }\left( \chi +\left| \Delta \right| \right)
+\pi k,  \label{PhasesNegDelta}
\end{equation}
for $\Delta >0$. For the opposite sign of the detuning, $\Delta
<0$ the mean photon numbers are given by the same Eq.(\ref
{Phot12Simplified}), while the phases read as

\begin{eqnarray}
\varphi _{10} &=&\frac{1}{2}Arc\sin \frac{1}{\varepsilon }\left(
\chi +\left| \Delta \right| \right) +\pi \left(
k+\frac{1}{2}\right) ,
\label{PhasesPosDelta} \\
\varphi _{20} &=&\frac{1}{2}Arc\sin \frac{1}{\varepsilon }\left(
\chi +\left| \Delta \right| \right) +\pi \left(
k-\frac{1}{2}\right) ,  \nonumber
\end{eqnarray}
($k=0,1,2,..$). Analyzing the system with help of a linear
treatment of quantum fluctuations in the $P$-representation we
have arrived \cite{adam1} to the following correlators
\begin{widetext}
\begin{eqnarray}
\left\langle{\delta n_{+} \choose \delta \varphi _{+}}\left(
\delta n_{+},\delta \varphi _{+}\right) \right\rangle
&=&\frac{1}{4\lambda n_{0}(\gamma +\lambda n_{0})}\left(
\begin{array}{cc}
4n_{0}\left[ \gamma (\gamma +\lambda n_{0})+(\chi -\mid \Delta
\mid )^{2}\right] , & -\lambda n_{0}(\chi -\mid \Delta \mid ) sign(\Delta ) \\
-2\lambda n_{0}(\chi -\mid \Delta \mid )sign(\Delta ), & -\lambda
\gamma
\end{array}
\right) ,  \label{dn+dfi+MeanFinal}\\
\nonumber\\
\left\langle {\delta n_{-} \choose \delta \varphi _{-}} \left(
\delta n_{-},\delta \varphi _{-}\right) \right\rangle
&=&\frac{1}{4\mid \Delta \mid \chi }\left(
\begin{array}{cc}
4n_{0}\chi (\chi -\mid \Delta \mid ), & 2\chi \gamma sign(\Delta ) \\
2\chi \gamma sign(\Delta ), & \frac{1}{n_{0}}(\gamma ^{2}-\mid
\Delta \mid (\chi -\mid \Delta \mid )
\end{array}
\right) ,  \label{dn-dfi-MeanFinal}
\end{eqnarray}
\end{widetext}
where $\delta n_{\pm }=\delta n_{2}\pm \delta n_{1}$, $\delta
\varphi _{\pm }=\delta \varphi _{2}\pm \delta \varphi _{1}$, and
$\delta n_{i}\left( t\right) =n_{i}\left( t\right) -n_{i0}$ and
$\delta \varphi _{i}\left( t\right) =\varphi _{i}\left( t\right)
-\varphi _{i0}$. These correlators is written in terms of the
stochastic variables, namely between photon-number sum and phase
sum in the modes, as well as between photon-number difference and
phase difference in the modes. What is now interesting for us is
the dependence of both photon number quantum correlation and phase
fluctuations on the wave plate parameter $\chi$.

\subsection{Photon-number quantum correlations in the presence of phase-locking}

Let us at first consider the quantum correlations in twin light
beams generated in self-phase locked NOPO. Intensity correlations
of twin light beams are usually characterized quantitatively by
the quantum fluctuations of the intensity difference between the
generated beams which is normalized to the corresponding shot
noise level. In this way the intense twin beams quantum
correlations have been experimentally observed several years ago
in NOPO operated above its threshold \cite{heidmann, mertz}. Up to
now the several interesting applications of intensity correlation
have been proposed and realized \cite{cao, laurat3, funk}. In this
section we analyze shortly the number-difference squeezing in the
presence of phase localizing process. Most published studies of
quantum correlations have usually been performed in the spectral
domain by measure of the corresponding squeezing spectra. However,
here we restrict ourselves by calculation only the integral
quantity. To characterize the photon-number correlation we address
to the variance of the fluctuations in the photon number
difference:
\begin{eqnarray}
R=\langle(a_{1}^\dag a_{1}-a_{2}^\dag a_{2})^{2}\rangle-(\langle
n_{1}\rangle-\langle n_{2}\rangle)^{2}.
\end{eqnarray}
This variance is expressed through the stochastic variables using
the relationship between normally-ordered moments of the operators
and the stochastic moments with respect to the $P$-function. In
the balance case $\langle n_{1}\rangle=\langle
n_{2}\rangle=\langle n\rangle$ and in the framework of the
stochastic variables we get

\begin{equation}
R=2\langle n\rangle+\langle \delta n_{-}^2\rangle
\label{R}.\end{equation} Then, using the formulas
(\ref{dn-dfi-MeanFinal}) we obtain

\begin{equation}
R=n_{0}\frac{|\Delta|+\chi}{|\Delta|}.
\end{equation}
Thus, the variance normalized to the level of fluctuations for the
coherent state is

\begin{equation}
R_{N}=\frac{R}{2n_0}=\frac{1}{2}(1+\frac{\chi}{|\Delta|}).
\end{equation}
For $\chi\rightarrow 0$, $R_N=1/2$, i.e. the normalized variance
reaches 50 percent relative to the quantum noise level of a
coherent state, in agreement with the result of the paper
\cite{kry1} devoted to an ordinary NOPO. As we see, for
$\chi/|\Delta|\ll1$ there is a small aggravation of the quantum
correlation between the photon numbers of the corresponding twin
fields in a self-phase locked NOPO in comparison with the case of
an ordinary NOPO.

The effect arising from the photon correlation in each modes as
well as between twin modes are usually investigated with the aid
of the second-order correlation functions

\begin{eqnarray}
g_{ii}=\langle a_{i}^{\dag2}a_{i}^{2} \rangle, \quad
g_{12}=\langle a_{1}^{\dag}a_{2}^{\dag}a_{1}a_{2} \rangle,
\end{eqnarray}
where ($i=1,2$). We calculate these correlations in the linear
treatment of quantum fluctuations and in terms of the stochastic
variables as

\begin{eqnarray}
&g_{11}=g_{22}=g=n_{0}^{2}+\frac{1}{4}(\langle \delta n_{+}^2 \rangle + \langle \delta n_{-}^2 \rangle), \\
\nonumber \ &g_{12}=n_{0}^{2}+\frac{1}{4}(\langle \delta n_{+}^2
\rangle - \langle \delta n_{-}^2 \rangle).
\end{eqnarray}
It is easy to verify that the following relation between the
correlation functions hold

\begin{equation}\label{g12}
g_{12}=g-\frac{1}{2}\langle \delta n_{-}^2
\rangle=g+\frac{n_{0}}{2}(1-\frac{\chi}{|\Delta|}),
\end{equation}

while the correlator $g$ equals to

\begin{equation}
g=n_{0}^{2}+\frac{\gamma}{4\lambda}(1+\frac{(\chi-|\Delta|)^2}{\gamma(\gamma+\lambda
n_0)})-\frac{n_{0}}{4}(1-\frac{\chi}{|\Delta|}).
\end{equation}
When the wave plate is not included, the formula (\ref{g12}) is
transformed to the relation between the correlation functions of
an ordinary NOPO \cite{kry1}.

As we see, the insertion of a polarization mixer slightly destroys
the photon number correlations produced by an ordinary NOPO.
Having discussed the influence of phase locking on photon-number
quantum correlations, we now turn our attention to investigate the
Wigner function of self-phase locked NOPO.

\subsection{Phase locking on the Wigner function}

It is clearly seen from the Eqs. (\ref{dn+dfi+MeanFinal}) and
Eqs.(\ref{dn-dfi-MeanFinal}) that for sufficiently small $\chi$
the terms $<\delta n_{+}^{2}>$, $<\delta n_{-}^{2}>$ and $<\delta
\varphi_{+}^{2}>$ does not depend on $\chi$, while the term
$<\delta \varphi_{-}^{2}>$ is inversely proportional to it.
Indeed, from Eqs.(\ref{dn-dfi-MeanFinal}) we have

\begin{equation}
\langle \delta\varphi_{-}^2\rangle=\frac{1}{4|\Delta|\chi
n_0}[\gamma^2-|\Delta|(\chi-|\Delta|)].
\end{equation}
Therefore, decreasing of $\chi$ in the correlator leads to
increasing of phase difference fluctuations, and at last with
$\chi=0$ we arrive to well known fact of phase diffusion in NOPO
without wave plate. This phenomenon also exhibits himself in the
framework of Wigner function and we turn to the consideration of
this point.

It has been found in \cite{adam1} from the general point of view
that the subharmonic modes has the phase space symmetry
properties: the Wigner functions $W_{1}$ and $W_{2}$ of the modes
have a two-fold symmetry under the rotation of the phase-space by
angle $\pi $ around its origin,

\begin{equation}
W_{i}\left( r,\theta \right) =W_{i}\left( r,\theta +\pi \right).
\label{WignerSymmetry}
\end{equation}
Here $r$, $\theta$ are the polar coordinates of the complex phase
space. We show below the results of numerical simulations in the
framwork of QSD method. Figure 1 plots the Wigner function of one
on the subharmonics of self-phase locked NOPO over transient time
and for all operational regimes. A clear self-phase locking in the
transition through the generation threshold is seen. The Wigner
function is almost Gaussian below threshold and is squeezed near
the threshold. Above-threshold the Wigner function has two
separated peaks due to the phase-localizing processes leading to
the phase locking of the subharmonic modes. The distance between
peaks increases with increasing the photon number (decreasing
$\lambda$). Note, that the decreasing of wave plate parameter
$\chi$ leads to less localization of these peaks (compare Fig.1(d)
($\chi/\gamma=0.5$) with Fig.1(c) ($\chi/\gamma=0.1$)).

\begin{figure}
\includegraphics[angle=270,width=0.22\textwidth]{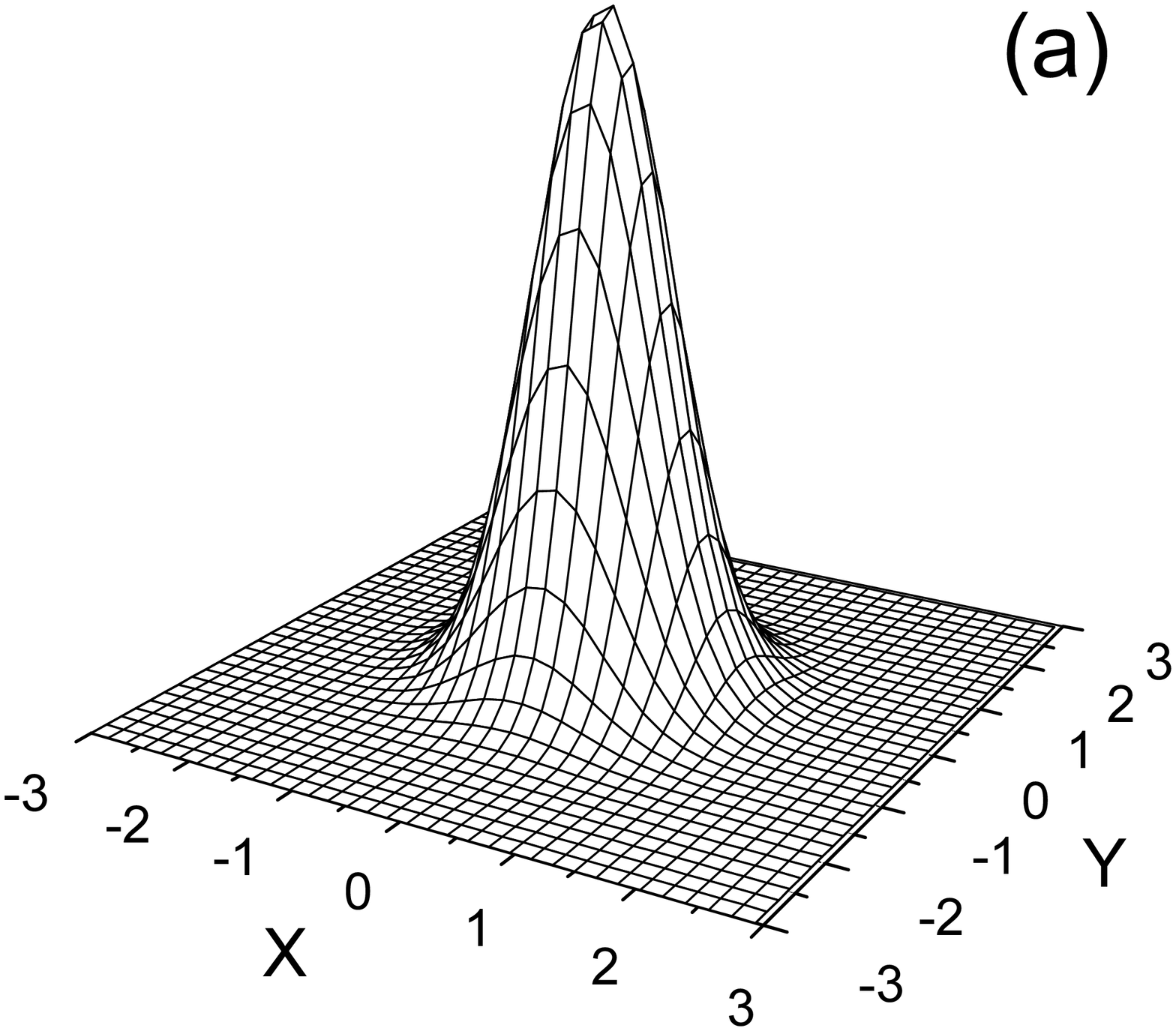}
\includegraphics[angle=270,width=0.22\textwidth]{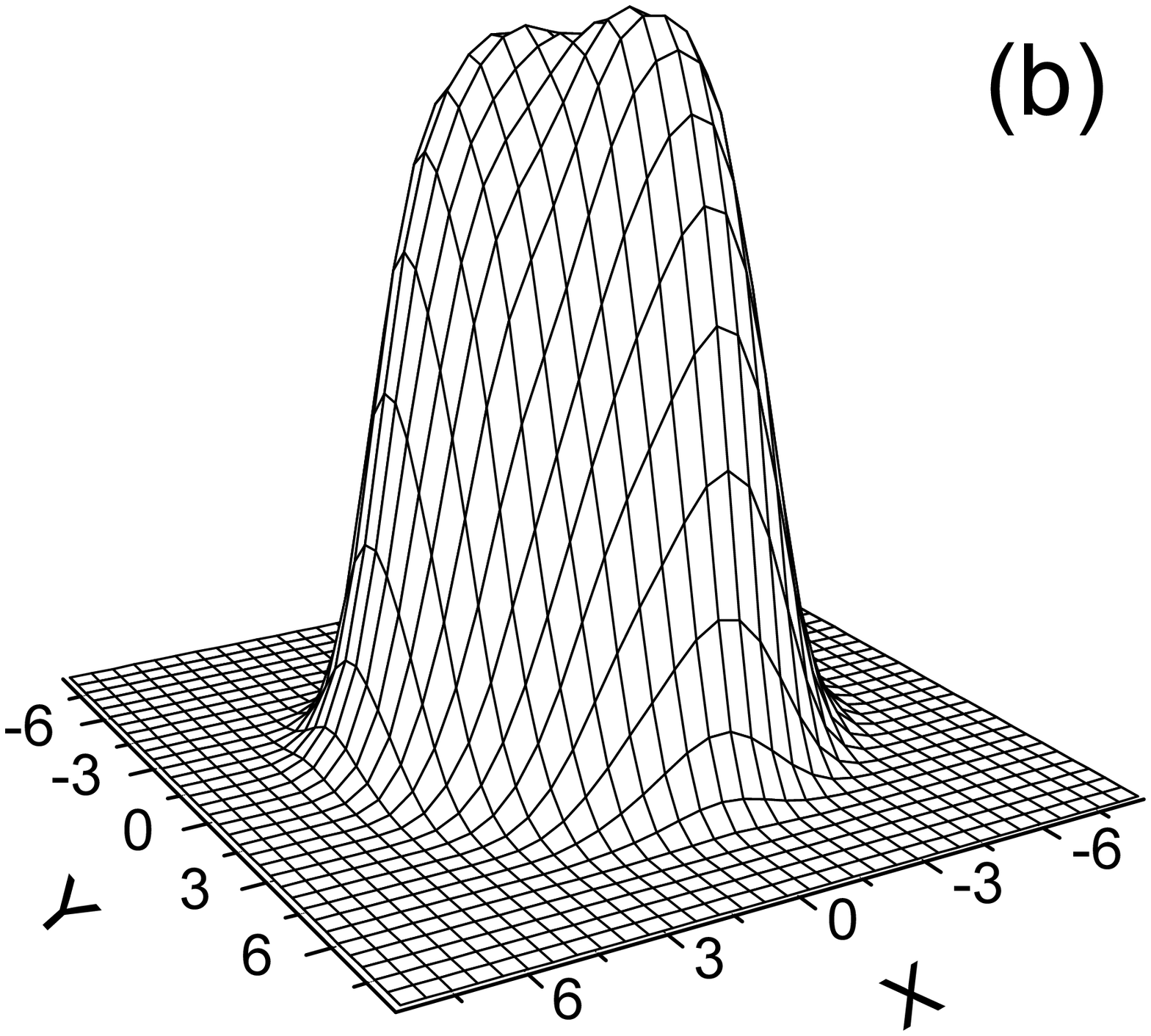}
\includegraphics[angle=270,width=0.22\textwidth]{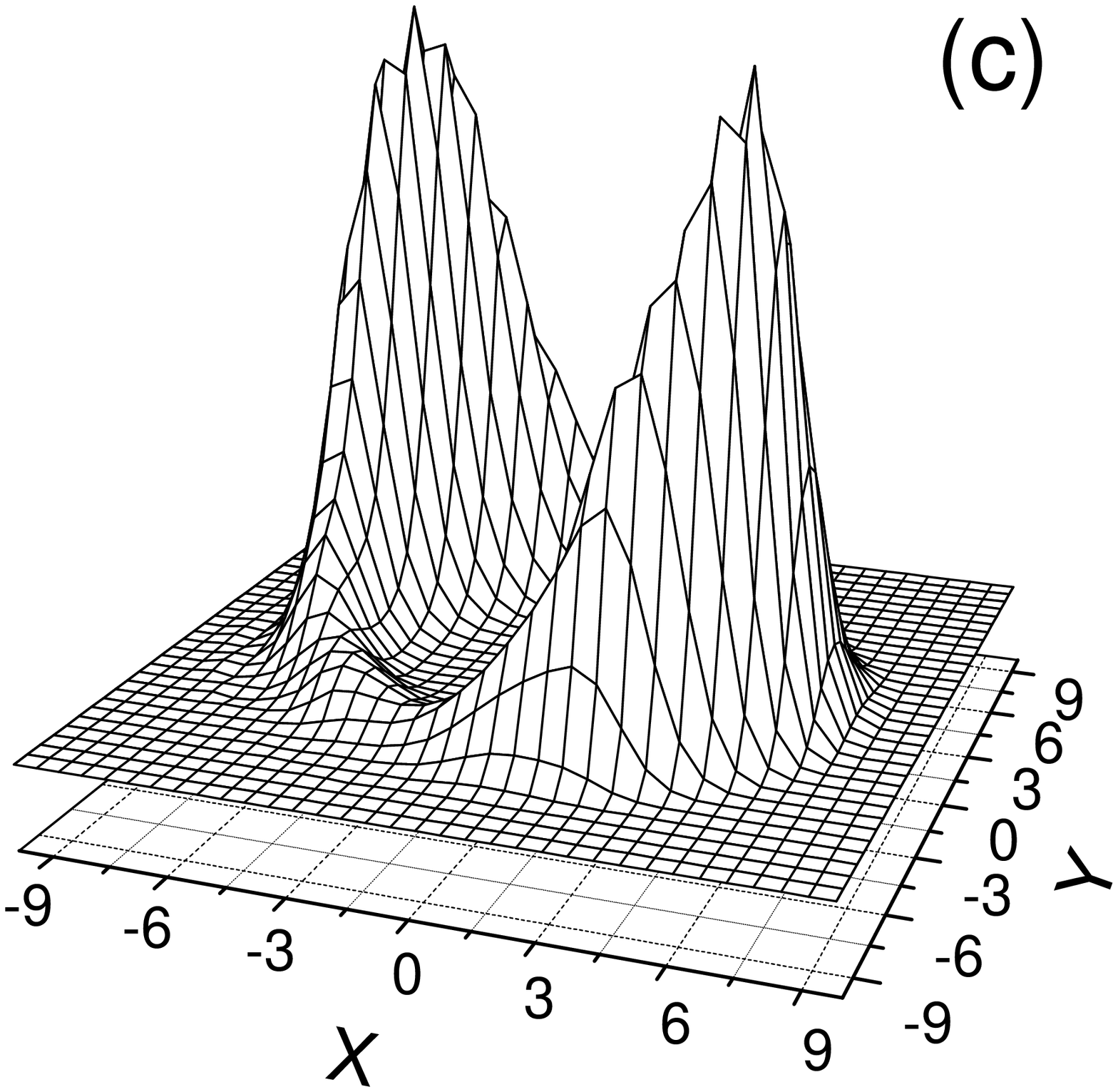}
\includegraphics[angle=270,width=0.22\textwidth]{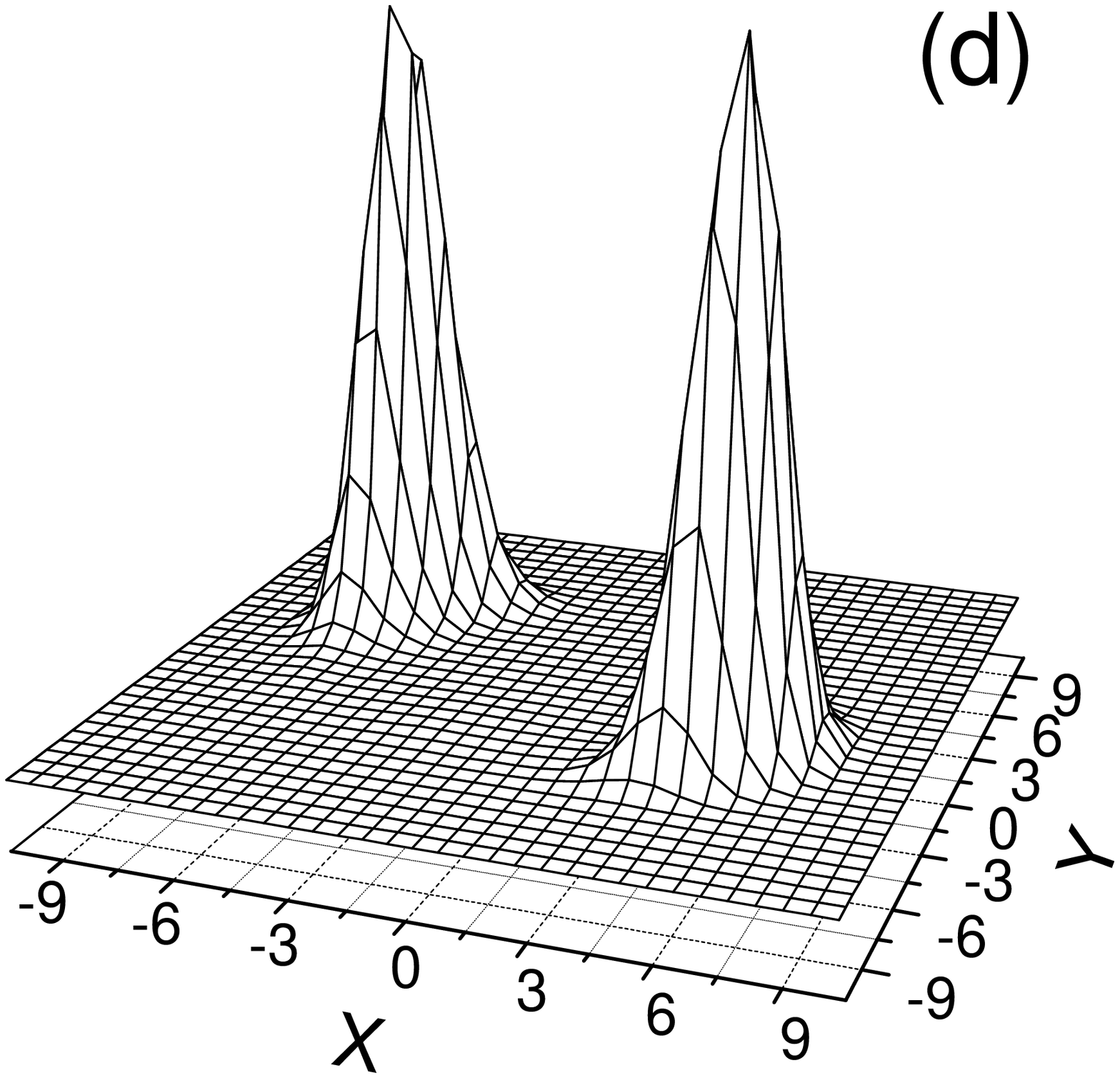}
\caption{ The Wigner function of the self-phase locked NOPO in the
stable regime for the parameters: $\lambda /\gamma =0.1$, $\Delta
_{1}/\gamma =\Delta _{2}/\gamma =10$, $\chi /\gamma =0.1$, a)
$\varepsilon /\gamma =5$, (below threshold), b) $\varepsilon
/\gamma =10$, (at threshold), c) $\varepsilon /\gamma =11$, (above
threshold), d) $\varepsilon /\gamma =11$, but $\chi /\gamma =0.5$.
Averaging is over 3000 quantum trajectories.} \label{Fig1}
\end{figure}

\section{Quadrature entanglement in the spectral domain}

In this section we continue investigation of CV entanglement in
the self-phase locked NOPO. In the previous studies \cite{adam1}
we have calculated the variances of the relevant distance
$V_{-}=V\left( X_{1}-X_{2}\right) $ and the total momentum
$V_{+}=V\left( Y_{1}+Y_{2}\right)$ of the quadrature amplitudes of
two modes $X_{k}=\frac{1}{\sqrt{2}}\left[ a_{k}^{+}\exp \left(
-i\theta _{k}\right) + a_{k}\exp \left( i\theta _{k}\right)
\right]$, $Y_{k}=\frac{i}{\sqrt{2}}\left[ a_{k}^{+}\exp \left(
-i\theta _{k}\right) - a_{k}\exp \left( i\theta _{k}\right)
\right]$, $\left(k=1,2\right)$, where $V(X)=\left\langle
X^{2}\right\rangle -\left\langle X\right\rangle ^{2}$ is a
denotation for the variance and $\theta _{k}$ is the phase of
local oscillator for the k-th mode. The two quadratures $X_{k}$
and $Y_{k}$ are non commuting observables. We have considered the
variance in the time-domain and have quantified the CV
entanglement by the inseparability criterion for the quantum state
of two optical modes \cite{duan}
\begin{equation}
V=\frac{1}{2}(V_{+}+V_{-})<1.  \label{EntanglementCriteria}
\end{equation}
It has been demonstrated recently that time-dependent quadrature
variance could be observed by means of time-resolved homodyne
measurements \cite{wenger}. This approach particularly allows for
applications in time-resolved quantum information protocols.
Nevertheless, so far squeezing as well as CV entanglement have
been mainly demonstrated in the spectral domain and not in the
time domain. Therefore, we calculate here the variance $V$ in the
spectral domain and in a fully quantum mechanical treatment of
self-phase locked NOPO below threshold, $E<E_{th}$.

We use the stochastic equations (\ref{a1StochEq}),
(\ref{b1StochEq}) linearized around the zero-amplitude solution
$\alpha_{i}^{0}=\beta_{i}^{0}=0$ in the following matrix form for
small fluctuations $\alpha_{i}(t)=\delta\alpha_{i}(t)$ and
$\beta_{i}(t)=\delta\beta_{i}(t)$, ($i=1,2$)

\begin{equation}
\frac{d\vec{L}}{dt} = -\hat{F}\vec{L} + \vec{R}.
\end{equation}
Here

\begin{equation} \label{eq:from_basics}
\hat{F} = \left( \begin{array}{ccc}
\hat{A} & -\hat{B} \\
-\hat{B}^* & \hat{A}^* \\
\end{array} \right),
\vec{R}\left(t\right) = \left( \begin{array}{ccc}
\vec{R}_\alpha\left(t\right) \\
\vec{R}_\beta\left(t\right) \\
\end{array} \right),\\
\vec{L}\left(t\right) = \left( \begin{array}{ccc}
\delta\vec{\alpha}(t) \\
\delta\vec{\beta}(t) \\
\end{array} \right)
\end{equation}
where $\delta \vec{\alpha} =\left( \delta \alpha _{1},\delta
\alpha _{2}\right) ^{T}$, $\delta \vec{\beta} =\left( \delta \beta
_{1},\delta \beta _{2}\right)^{T}$, $\vec{R}_{\alpha}
=\left(R_{1}, R_{2}\right)^{T}$, $\vec{R}_{\beta}
=\left(R^{+}_{1},R^{+}_{2}\right)^{T}$ are two-dimensional column
vectors. The $4\times4$ matrix $\hat{F}$ is written in the block
form with $2\times2$ matrices

\begin{equation}
\hat{A}=\left(
\begin{array}{cc}
\gamma +i\Delta & i\chi \\
i\chi & \gamma +i\Delta
\end{array}
\right) ,\;\;\hat{B}=\varepsilon \hat{\sigma},\;\; \hat{\sigma} =
\left(
\begin{array}{cc}
0 & 1 \\
1 & 0
\end{array}
\right).  \label{eq:AandBMatrices}
\end{equation}
The noise correlators are determined as

\begin{eqnarray} \label{eq:shumer}
\langle\vec{R}_\alpha(t)\vec{R}^T_\alpha(t')\rangle =
\hat{D}\delta(t - t')
\end{eqnarray}
with the following diffusion matrix:

\begin{equation} \label{eq:DMatrix}
\hat{D} = \left( \begin{array}{ccc}
\hat{B} & 0 \\
0 & \hat{B}^* \\
\end{array} \right).
\end{equation}
The correlation functions of the quantum fluctuations are obtained
in the following form (we show results of \cite{adam1} with
changing some misprints)

\begin{widetext}
\begin{eqnarray}
\left\langle \delta \alpha \left( \delta \alpha \right)
^{T}\right\rangle &=&\frac{\varepsilon }{2\left(S^{4}-4\Delta
^{2}\chi ^{2}\right)}\left[ \gamma \left(
\begin{array}{cc}
-2\chi \Delta, & S^{2} \\
S^{2}, & -2\chi \Delta
\end{array}
\right) -i\left(
\begin{array}{cc}
\chi \left(S^{2} -  2\Delta ^{2}\right), & \Delta \left(S^{2} -  2\chi ^{2}\right) \\
\Delta \left(S^{2} -  2\chi ^{2}\right), & \chi \left(S^{2} -
2\Delta ^{2}\right)
\end{array}
\right) \right],  \nonumber \\
 \nonumber \\
\left\langle \delta \alpha \left( \delta \beta \right) ^{T
}\right\rangle &=&\frac{\varepsilon ^{2}}{2\left(S^{4}-4\Delta
^{2}\chi ^{2}\right)}\left(
\begin{array}{cc}
S^{2}, & -2\chi \Delta \\
-2\chi \Delta, & S^{2}
\end{array}
\right) , \label{eq:dadbMeanFinal}
\end{eqnarray}
\end{widetext}
where $S^{2}$ is introduced as

\begin{equation}
S^{2}=\gamma ^{2}+\chi ^{2}+\Delta ^{2}-\varepsilon ^{2}.
\label{eq:Determinant}
\end{equation}
These correlators are obtained in a steady state regime and do not
dependent on time. However, for spectrally resolved measurements
of the cavity output fields we need in the correlation functions
in the spectral domain. To calculate them we use the
Fourier-transformed linearized equations for the stochastic
amplitudes

\begin{widetext}
\begin{eqnarray}
\vec{\alpha}(t) =
\frac{1}{\sqrt{2\pi}}\int_{-\infty}^{\infty}e^{\imath\omega
t}\vec{\alpha}(\omega) d\omega, ~~~~~\vec{\alpha}(\omega) =
\frac{1}{\sqrt{2\pi}}\int_{-\infty}^{\infty}e^{-\imath\omega
t}\vec{\alpha}(t)dt,\\
\vec{\beta}(t) =
\frac{1}{\sqrt{2\pi}}\int_{-\infty}^{\infty}e^{\imath\omega
t}\vec{\beta}(\omega) d\omega, ~~~~~\vec{\beta}(\omega) =
\frac{1}{\sqrt{2\pi}}\int_{-\infty}^{\infty}e^{-\imath\omega
t}\vec{\beta}(t)dt,
\end{eqnarray}
\end{widetext}
which are obtained as given below
\begin{equation} \label{eq:main_eqation_fourier}
\left\{ \begin{array}{ll} \imath\omega\delta\vec{\alpha}(\omega) =
-\hat{A}\delta\vec{\alpha}(\omega)
+\hat{B}\delta\vec{\beta}(\omega)
+\vec{R}_\alpha(\omega), \\
\imath\omega\delta\vec{\beta}(\omega) =
\hat{B}^*\delta\vec{\alpha}(\omega)
-\hat{A}^*\delta\vec{\beta}(\omega) +\vec{R}_\beta(\omega).
\end{array} \right.
\end{equation}
Here, the nonzero correlation functions of noise terms in the
frequency space are
\begin{eqnarray}
\langle\vec{R}\left(\omega\right)\vec{R}^T
\left(\omega'\right)\rangle = \hat{D}\delta\left(\omega +
\omega'\right).\label{Rcorrelation}
\end{eqnarray}
We rewrite these equations in the more compact form for
$\vec{L}(\omega)=\left( \begin{array}{ccc}
\delta\vec{\alpha}(\omega) \\
\delta\vec{\beta}(\omega) \\
\end{array} \right)$ as
\begin{equation}
\vec{L}(\omega) =(\hat{F}+\imath\omega\hat{I})^{-1}\vec{R}(\omega)
\end{equation}
and then using the correlators from Eq.((\ref{Rcorrelation})), we
explicitly calculate

\begin{widetext}
\begin{eqnarray}
\langle\vec{L}(\omega)\vec{L}^T(\omega')\rangle =
\langle(\hat{F}+\imath\omega\hat{I})^{-1}\vec{R}(\omega)\vec{R}^T(\omega')(\hat{F}+\imath\omega'\hat{I})^{-1T}\rangle
=
\nonumber\\
(\hat{F}+\imath\omega\hat{I})^{-1}\hat{D}(\hat{F}+\imath\omega'\hat{I})^{-1T}\delta(\omega
+ \omega').
\end{eqnarray}
\end{widetext}
This result can be transformed to the following form

\begin{eqnarray}
\langle\vec{L}(\omega)\vec{L}^T(\omega')\rangle &=&
(\hat{F}+\imath\omega\hat{I})^{-1}(\hat{F}+\imath\omega'\hat{I})^{-1}\hat{D}\delta(\omega
+ \omega') =
\nonumber\\
&&=(\hat{F^2}+\omega^2\hat{I})^{-1}\hat{D}\delta(\omega +
\omega')\label{eq:corrMatrixForm}
\end{eqnarray}
by using the formula

\begin{equation}
\hat{D}(\hat{F}+\imath\omega\hat{I})^{-1T} =
(\hat{F}+\imath\omega\hat{I})^{-1}\hat{D}.
\end{equation}
The further simplification of this result is connected with the
structures of the matrices $\hat{F}$, $\hat{D}$ which are written
in the block forms (\ref{eq:from_basics}),(\ref{eq:DMatrix}). As a
consequence we have

\begin{figure}
\includegraphics[angle=0,width=0.35\textwidth]{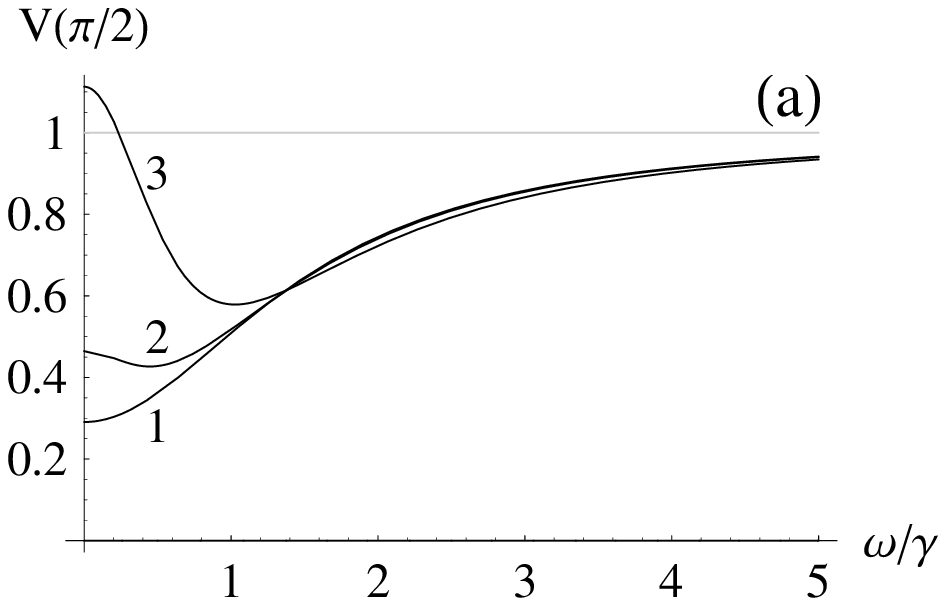}
\includegraphics[angle=0,width=0.35\textwidth]{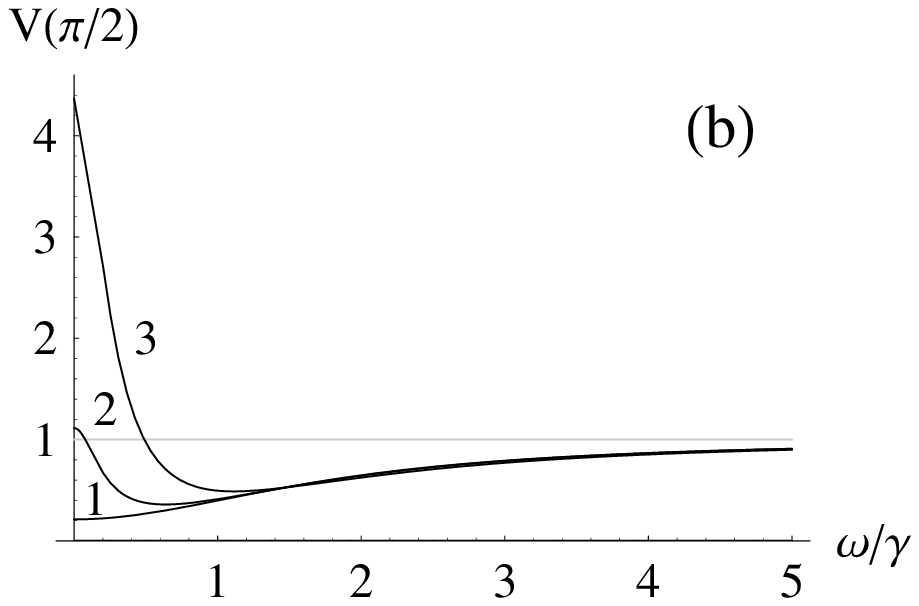}
\includegraphics[angle=0,width=0.35\textwidth]{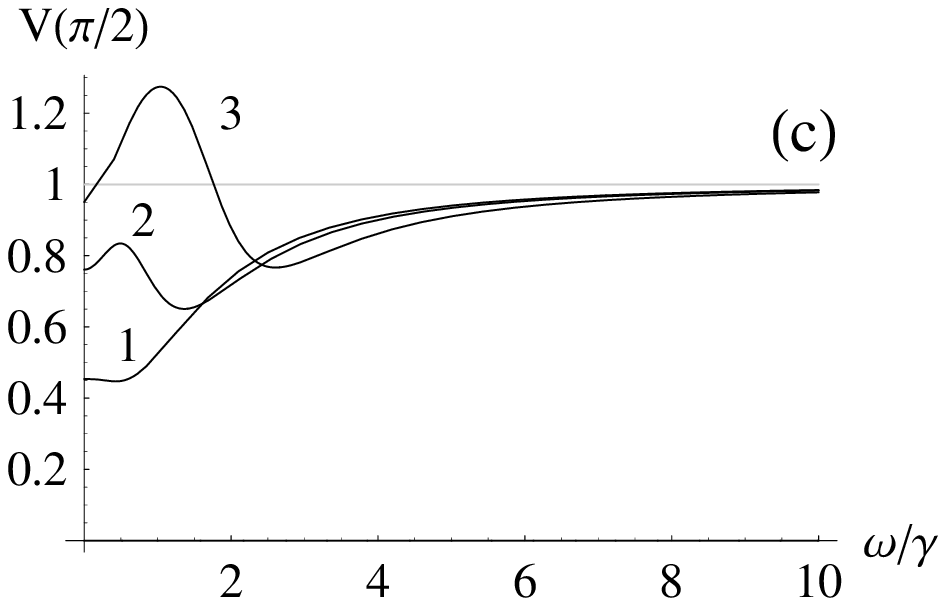}
\caption{The spectral variances versus $\omega/\gamma$. The
parameters are: a)$\Delta/\gamma =0$, $\chi /\gamma =0, 0.2, 0.5$
(curve 1, 2, 3), $\varepsilon /\varepsilon_{th} =0,5$.
b)$\Delta/\gamma =0$, $\chi /\gamma =0, 0.2, 0.5$(curve 1, 2, 3),
$\varepsilon /\varepsilon_{th} =0,8$. c)$\Delta/\gamma =0.2, 0.5,
1$(curve 1, 2, 3), $\chi /\gamma =0.05$, $\varepsilon
/\varepsilon_{th} =0,5$} \label{Fig2}
\end{figure}

\begin{eqnarray} \label{eq:FExpanded}
\hat{F}^2 + \hat{I}\omega^2 = \left(
\begin{array}{lll}
a\hat{I} + b \hat{\sigma} & -c \hat{\sigma} \\
-c^* \hat{\sigma} & a^*\hat{I} + b^* \hat{\sigma}
\end{array}
\right),
\end{eqnarray}
where $I$ is the identity matrix and the following notations are
used

\begin{eqnarray}
a&=& (\gamma + i \Delta)^2 - \chi^2 + \varepsilon^2 + \omega^2, \\
\nonumber b &=& 2 i \chi(\gamma + i \Delta),\\ \nonumber c &=& 2
\gamma \varepsilon.
\end{eqnarray}
As a result, the correlation matrices (\ref{eq:corrMatrixForm})
can be also written in the following block form

\begin{equation} \label{eq:resultFind}
\langle\vec{L}(\omega)\vec{L}^T(\omega')\rangle = \left(
\begin{array}{ccc}
\hat{G}_0 & \hat{G}_1 \\
\hat{G}_1^* & \hat{G}_0^* \\
\end{array} \right)\delta(\omega + \omega'),
\end{equation}
where the matrices $\hat{G}_0$, $\hat{G}_1$ can be calculated from
the equations

\begin{eqnarray}
\left(
\begin{array}{lll}
a\hat{I} + b \hat{\sigma} & -c \hat{\sigma} \\
-c^* \hat{\sigma} & a^*\hat{I} + b^* \hat{\sigma}
\end{array}
\right) \left(
\begin{array}{ccc}
\hat{G}_0 & \hat{G}_1 \\
\hat{G}_1^* & \hat{G}_0^* \\
\end{array}
\right) = \left(
\begin{array}{lll}
\varepsilon \hat{\sigma} & 0 \\
0 & \varepsilon \hat{\sigma}
\end{array}
\right),
\end{eqnarray}

which can be transformed to the following form:

\begin{eqnarray}
\left\{
\begin{array}{ccc}
 a \hat{G}_0 + b \hat{\sigma}\hat{G}_0 - c \hat{\sigma}\hat{G}_1^* = \varepsilon \hat{\sigma}\\
 a \hat{G}_1 + b \hat{\sigma}\hat{G}_1 - c \hat{\sigma}\hat{G}_0^* = 0. \\
\end{array}
\right.
\end{eqnarray}
The solution of these equations is obtained as

\begin{widetext}
\begin{eqnarray}
\hat{G}_0 = \frac{\varepsilon}{d^2 - e^2}\big( (db^* -
ea^*)\hat{I} + (da^* - eb^*)\hat{\sigma}\big), \hat{G}_1 =
\frac{\varepsilon c }{d^2 - e^2}(d\hat{I} - e \hat{\sigma}),
\end{eqnarray}
\end{widetext}
where $d = aa^* + bb^* - cc^*$ and $e = ab^* +
ba^*$.

The further calculation of the spectral variance using the
$P$-representation is standard. A detailed description of the
method can be found in \cite{drummond}. We include also the output
coupler transmissivity $2\gamma$ for cavity output fields so that
the frequency spectrum of the squeezing variance
$V(\theta,\omega)$ corresponding to the integral variance in the
formula (\ref{EntanglementCriteria}) is calculated to be
\begin{eqnarray}
V(\theta,\omega) = 1 + S^{(1)}(\theta,\omega-\Delta) +
S^{(1)}(\theta,-\omega-\Delta),
\end{eqnarray}
where
\begin{eqnarray}
S^{(1)} = \frac{\gamma}{\pi}(G_{1,11}(\omega) + G_{1,22}(-\omega)
 \\ \nonumber +2 Re[e^{2 i \theta}G_{0,21}^*(\omega)]),
\end{eqnarray}
and  $G_{0,21}^*$, $G_{1,11}$, $G_{1,22}$ are the elements of the
matrices $\hat{G}_0^*$, $\hat{G}_1$. The variance
$V(\theta,\omega)$ of the output quadrature amplitudes is
symmetric in frequency around $\omega=0$, and equal to unity in
the vacuum or coherent signal case, and can only reach its minimum
value at the definite frequency. For the case of ordinary NOPO, if
$\chi=0$, this result is coincided with the analogous one well
known result \cite{drummond}. Typical results for self-locked NOPO
are shown in Figs.2, for the various parameters $\chi$ and for
$\theta=\pi/2$ chosen to minimize the noise level
$V(\theta,\omega)$. The standard quantum limit is shown in grey.
We see that the minimal variance spectra remain less than unity
for all frequencies only for the cases of small values of the
parameter $\chi$.

Figures \ref{Fig2}(a), \ref{Fig2}(b) plot the squeezing spectra in
the absence of the detunings $(\Delta_1 = \Delta_2 = 0)$ for two
values of the pump field: $\varepsilon = 0.5\varepsilon_{th}$
(Fig. \ref{Fig2}(a)) and $\varepsilon = 0.8\varepsilon_{th}$ (Fig.
\ref{Fig2}(b)), and for the various values of the parameter
$\chi$: $\chi=0$ (curves 1), $\chi=0.2$ (curves 2), $\chi = 0.5$
(curves 3). The excellent squeezing spectra centered at $\omega
=0$, occurs near threshold for $\chi = 0$ that is the case of an
ordinary NOPO \cite{drummond}. We point out that there is a
corresponding decrease in the squeezing as the parameter $\chi$
increases. Note, that for phase-locked NOPO the minima of spectra
are shifted form the zero frequency (see Fig.2 (a), (b) curves 2
and 3) even in the case of zero detunings. Figures \ref{Fig2}(c)
plot squeezing spectra for the case of nonzero detuning. These
results for $\chi = 0$ are in a good agreement with the results
obtained for an ordinary NOPO \cite{drummond}.

\section{Self-pulsing regime and entanglement}

Now let us pay our attention to the case of classically
nonstationary regime of generation, when the inequality
(\ref{SolutionCondition}) does not valid. In this regime
analytical treatment of semiclassical and quantum equations is
complicated, therefore, we simulate Eqs. (\ref{a1StochEq}) and
(\ref{b1StochEq}) on one side, and use the QSD method to obtain
numerical solution of Eq. (\ref{MasterEq}) on the other side. We
find that the semiclassical solution of the equation
(\ref{a1StochEq}), (\ref{b1StochEq}), without the noise terms and
for $\beta_i=\alpha_i^*$, exhibits the self-pulsing instability:
the photon number of intracavity modes oscillates periodically. We
demonstrate this oscillations in Fig.3 (a) (curve 1) for one of
the modes. It appears that self-pulsing exists in the whole range
of violation of the inequality (\ref{SolutionCondition}), and does
not depend on the what parameter is changed to violate it. We
present also the phase space trajectory of the semiclassical
solution in Fig.4 (a), which has the form of squeezed circle. The
individual quantum trajectory of the single mode of self-locked
NOPO for the same parameters as in Fig.3(a) is presented in Fig.3
(b) as a quantum mechanical calculation on the base of QSD method.
It is seen that trajectory repeats the oscillations of
semiclassical solution.

\begin{figure}
\includegraphics[angle=0,width=0.35\textwidth]{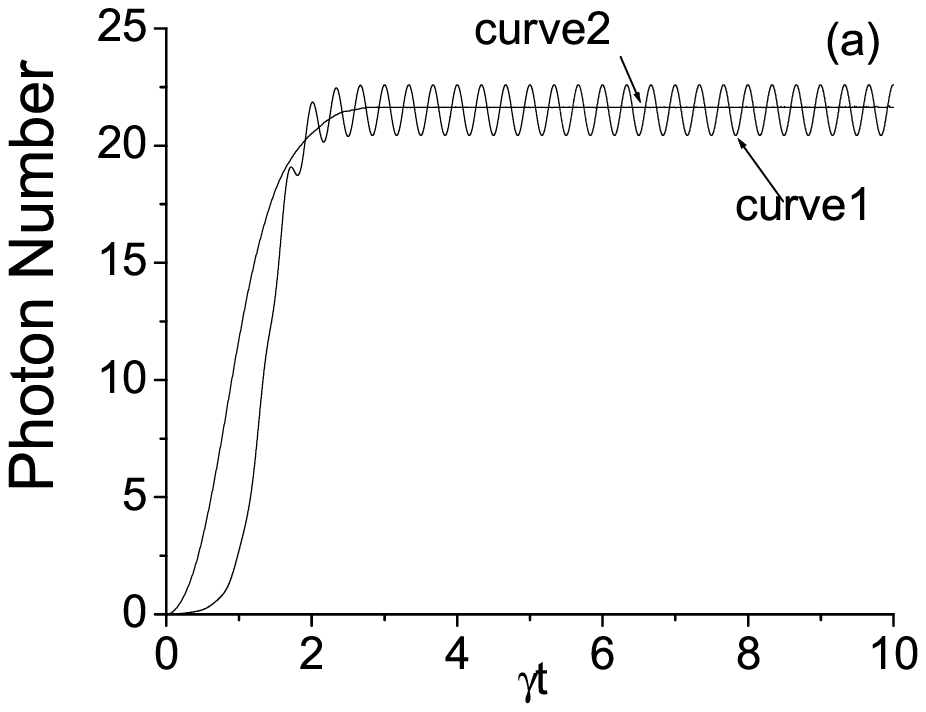}
\includegraphics[angle=0,width=0.35\textwidth]{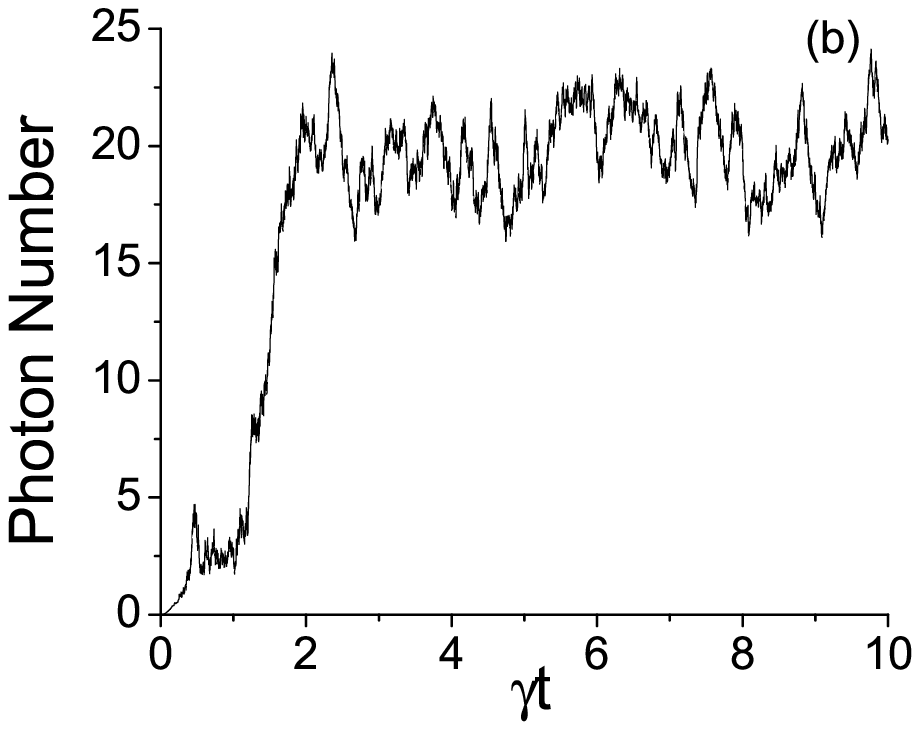}
\caption{Time dependence of the photon number: (a) classical
trajectory (curve 1) and quantum ensemble averaged result (curve
2), and (b) quantum  trajectory of self-phase locked NOPO in the
regime of self-pulsing. The parameters are: $\lambda /\gamma
=0.1$, $\Delta
_{1}/\gamma =10$, $\Delta _{2}/\gamma =-5$, $\chi /\gamma =0.5$,  $%
\varepsilon /\gamma =4$. Curve 2 on the Fig.3(a) involve averaging
over 12000 quantum trajectories.} \label{Fig3}
\end{figure}

\begin{figure}
\includegraphics[angle=0,width=0.35\textwidth]{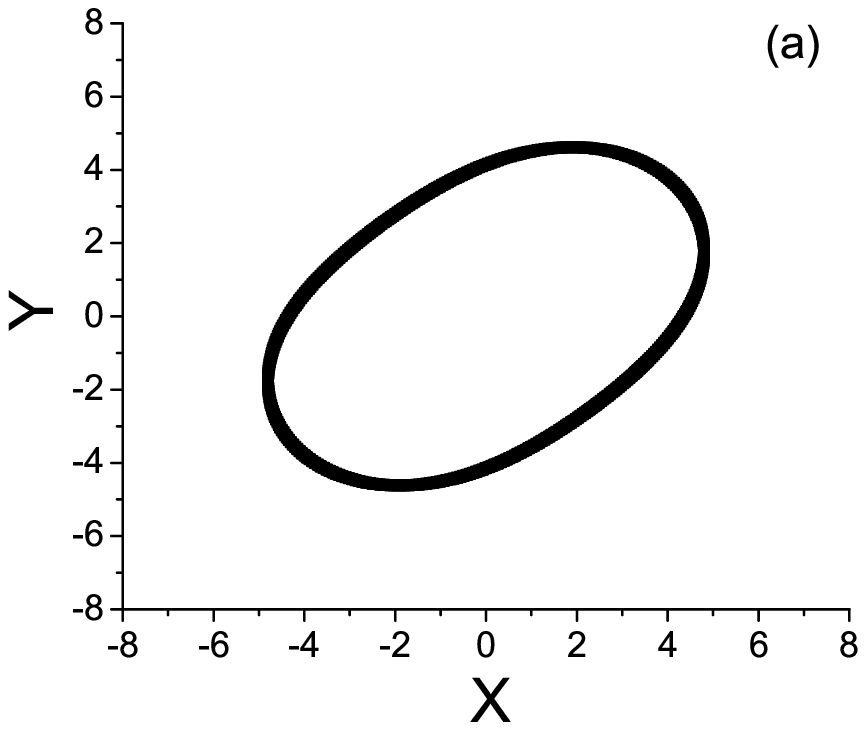}
\includegraphics[angle=0,width=0.37\textwidth]{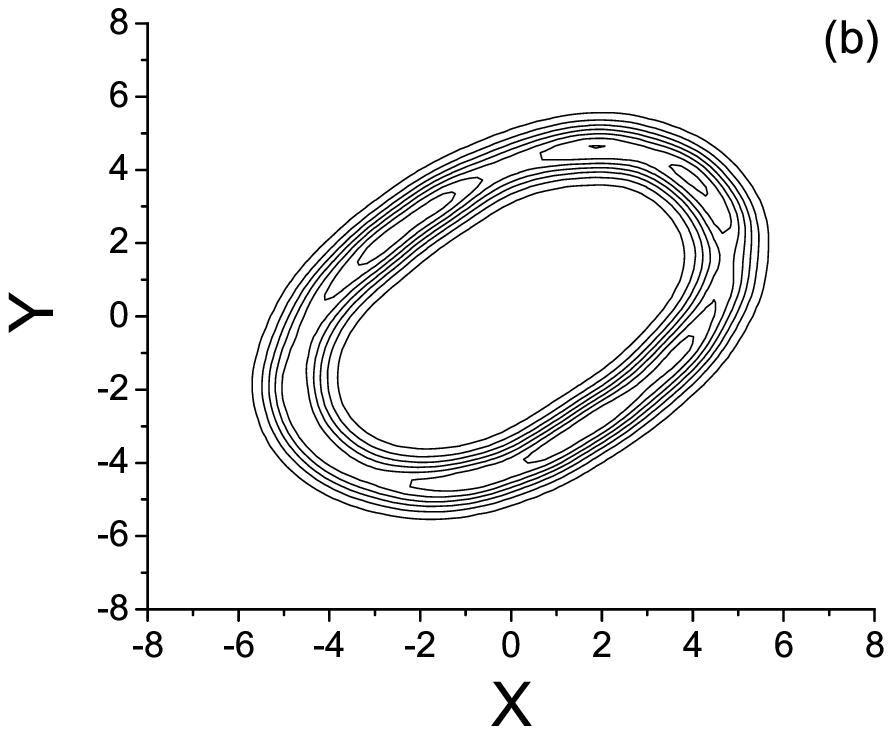}
\caption{Semiclassical phase space trajectory (a) and
contour-plots of the Wigner function (b) of the self-phase locked
NOPO in the regime of self-pulsing.
The parameters are: $\lambda /\gamma =0.1$, $\Delta _{1}/\gamma =0.1$, $%
\Delta _{2}/\gamma =-0.1$, $\chi /\gamma =0.5$, $\varepsilon
/\gamma =3$. Averaging in Fig.4 (b)is over 3000 quantum
trajectories.} \label{Fig4}
\end{figure}

Due to the quantum noise, the averaged over ensemble of quantum
trajectories result does not exhibit oscillations (see Fig.3(a),
curve 2). The analogous situation takes place for various quantum
systems, particularly, also for chaotic systems, where quantum
trajectory reflects the chaotic nonstationary behavior of the
semiclassical trajectory, but ensemble averaged results has
stationary solution (see \cite{manvelyan}). It has been shown in
Ref.\cite{manvelyan}, that quantum chaos manifests itself,
particularly, in the Wigner function: it repeats the shape of
Poincar\'{e} section of semiclassical counterpart. So, it is
expected that \emph{the features of the} phase trajectory
(Fig.4(a)) will be reflected in the Wigner function also for the
system under consideration. Indeed, as it is seen from Fig.4(a)
and Fig.4(b), where the contour-plots of the Wigner function is
depicted, it is really the case. Note, that the result of Fig.4(b)
is obtained in the framework of QSD method. Thus, we can conclude
that the photon number of self locked NOPO is stationary in the
full quantum treatment, even when the semiclassical counterpart in
nonstationary. However, self pulsing manifests itself in the
Wigner function, which is rather different from the case of
stationary stable regime (see Fig.1 and Fig. 4(b)).

\begin{figure}
\includegraphics[angle=270,width=0.35\textwidth]{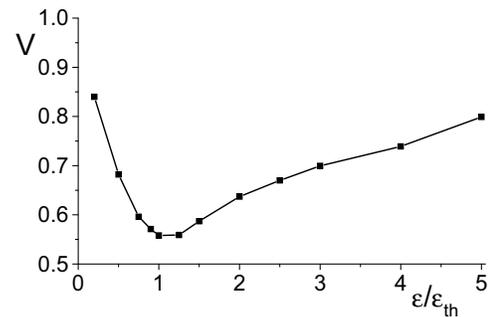}
\caption{Entanglement versus pump amplitude. The parameters are: $%
\lambda /\gamma =0.1$, $\Delta _{1}/\gamma =10$, $\Delta _{2}/\gamma =-10$, $%
\chi /\gamma =0.1$.} \label{Fig5}
\end{figure}
\begin{figure}
\includegraphics[angle=270,width=0.35\textwidth]{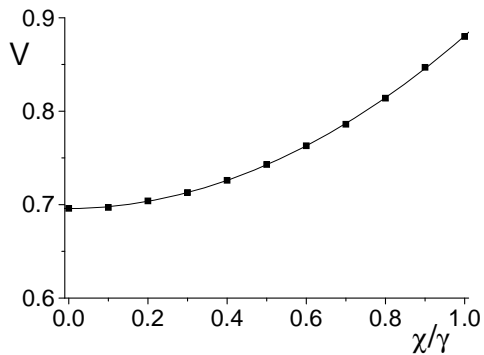}
\caption{Entanglement versus wave plate parameter $\chi /\gamma$
(squares - numerical result, line - quadratic fit). The parameters
are: $\lambda /\gamma =0.1$, $\Delta _{1}/\gamma =10$, $\Delta
_{2}/\gamma =-10$, $\varepsilon /\gamma =3$.} \label{Fig6}
\end{figure}

It is also interesting to compare the result of Fig.4(b), which
has a slightly squeezed circle form, with the Wigner function of
ordinary NOPO ($\chi=0$), presented in the paper
\cite{kheruntsyan}, which has the form of circle. This form is the
manifestation of phase diffusion. Indeed, the phase in phase space
is proportional to the angle between the $Ox$ axe and
radius-vector of the point. As the Wigner function distributed
uniformly to all directions, this mean that the phase operator is
entirely uncertain, which is equivalent to phase diffusion. For
the case of self locked NOPO the distribution is almost uniform to
all directions, which means that the modified phase diffusion also
takes place here. The squeezed form of the circle is due to
oscillation of photon number.

Analogous results, but for triply resonant cavity have been
obtained numerically by Gro$\beta$ and Boller \cite{Boller} in the
framework of semiclassical treatment: the phases of the modes of
self locked NOPO in the self-pulsing regime diverge.  Moreover,
the rate of the phase diffusion is the same as for an ordinary
NOPO. It is interesting that the semiclassical phase space
trajectory in Fig.4(a) has similar form as in the case of triply
resonant NOPO.

Another interesting question, which arises from studying
nonstationary regime is: whether entanglement takes place in this
regime or not? And, if yes, does it reach the same extent or
perhaps is higher with respect to the stationary regime.
Nevertheless, we understand that it will not have the desirable
property of self-phase locking. For answering these questions, we
consider the dependence of dispersion $V$
(\ref{EntanglementCriteria}) on both driving field amplitude in
units of its threshold value $\varepsilon/\varepsilon_{th}$, and
quarter wave plate parameter $\chi$. At first, we numerically
estimate that $\varepsilon_{th}/\gamma=1$ for the parameters used,
and then we consider the dependence of $V$ on
$\varepsilon/\varepsilon_{th}$. Then, we perform calculations and
the results are presented in Fig.5. It is seen, that the variance
drops down near the threshold and achieve the minimal value
$V=0.55$ at the threshold. Then, it increases by increasing the
driving amplitude. This behavior is similar to that for the case
of stable operational regime of self-phase locked NOPO
\cite{adam1}.

We also present here the dependence of $V$ versus wave plate
parameter $\chi $ in Fig. 6, where the decreasing of entanglement
degree by increasing of $\chi$ is clearly evident. This means,
that while, as mentioned in Sec.II, the localization of phases of
the modes is improved with increasing of $\chi$, the entanglement
is worsening.

It is remarkable, that this dependence is exactly quadratic: we
fit the numerical results (squares in Fig.6) with quadratic curve
and find excellent coincidence (solid curve in Fig.6).

\section{Conclusion}

In conclusion, we have studied in the full quantum mechanical
manner the properties of the light beams generated in self-phase
locked NOPO. We have continued the recent investigations of the
quantum aspects of this device in the steady state, stable regime
of generation \cite{adam1} in one side and also we have considered
the specific signatures of the NOPO containing a birefringent
element in the unstable regime of generation on the other side. In
this latter case, the system exhibits the both self-pulsing
temporal behavior and a new type of phase diffusion as it has been
demonstrated by numerical simulations on the framework of the
Wigner function. The entanglement does occur in this regime too,
but suffer some worsening: it decreases quadratically when wave
plate parameter increases linearly. Considering the stable regime
of generation, we conclude that significant noise reduction in
intensity difference variance as well as in the quadrature
squeezing spectra are possible for small parameters $\chi / \gamma
\ll 1$. A detailed analysis of two-mode squeezing spectra for
self-phase locked NOPO below threshold has been given, using a
P-presentation in applications to the experiment recently
performed \cite{laurat1}. We have also analyzed photon-number
correlations in the presence of phase localization which has also
been studied on framework of the Wigner function.

\section{Acknowledgements}

This work has been supported by INTAS Grant $No$ 04-77-7289 and
ANSEF Grant $No$ 05-PS-compsci-89-66.

\end{document}